\pdfoutput=1
\documentclass[%
  aps,
  prb,
  citeautoscript,
  superscriptaddress,
  amsmath,
  amssymb,
  reprint
]{revtex4-1}

\usepackage[utf8]{inputenc}
\usepackage[T1]{fontenc}
\usepackage{graphicx}
\usepackage{glossaries}
\usepackage{mathtools}
\usepackage[version=4]{mhchem}
\usepackage{textcomp}
\usepackage{units}
\usepackage{braket}

\usepackage{xpatch}
\makeatletter
\xpatchcmd{\@ssect@ltx}{\@xsect}{\protected@edef\@currentlabelname{#8}\@xsect}{}{}
\xpatchcmd{\@sect@ltx}{\@xsect}{\protected@edef\@currentlabelname{#8}\@xsect}{}{}
\makeatother

\usepackage[colorlinks=true,urlcolor=blue,linkcolor=blue,citecolor=blue]{hyperref}

\newcommand{\ie}{\emph{i.e.}}
\newcommand{\eg}{\emph{e.g.}}
\newcommand{\V}[1]{\boldsymbol #1}
\newcommand{\dee}{\mathrm{d}}
\renewcommand{\Re}{\mathrm{Re}}
\renewcommand{\Im}{\mathrm{Im}}
\newcommand{\angstrom}{\mbox{\normalfont\AA}}

\glsdisablehyper
\newacronym{dft}{DFT}{density-functional theory}
\newacronym{hc}{HC}{hot carrier}
\newacronym{ks}{KS}{Kohn--Sham}
\newacronym{lcao}{LCAO}{linear combination of atomic orbitals}
\newacronym{rt}{RT}{real-time}
\newacronym{lspr}{LSPR}{localized surface plasmon resonance}
\newacronym{np}{NP}{nanoparticle}
\newacronym{rpa}{RPA}{random-phase approximation}
\newacronym{tddft}{TDDFT}{time-dependent density-functional theory}
\newacronym{xc}{XC}{exchange-correlation}

\begin{document}

\title{Hot-Carrier Generation in Plasmonic Nanoparticles:\\ The Importance of Atomic Structure}

\date{\today}

\author{Tuomas\ P.\ Rossi}
\email{tuomas.rossi@chalmers.se}
\author{Paul\ Erhart}
\email{erhart@chalmers.se}
\affiliation{Department of Physics, Chalmers University of Technology, SE-412~96 Gothenburg, Sweden}

\author{Mikael\ Kuisma}
\email{mikael.j.kuisma@jyu.fi}
\affiliation{Department of Chemistry, Nanoscience Center, University of Jyv\"askyl\"a, FI-40014 Jyv\"askyl\"a, Finland}

\begin{abstract}
Metal nanoparticles are attractive for plasmon-enhanced generation of hot carriers, which may be harnessed in photochemical reactions.
In this work, we analyze the coherent femtosecond dynamics of photon absorption, plasmon formation, and subsequent hot-carrier generation through plasmon dephasing using first-principles simulations.
We predict the energetic and spatial hot-carrier distributions in small metal nanoparticles and show that the distribution of hot electrons is very sensitive to the local structure.
Our results show that surface sites exhibit enhanced hot-electron generation in comparison to the bulk of the nanoparticle.
While the details of the distribution depend on particle size and shape, as a general trend lower-coordinated surface sites such as corners, edges, and \{100\} facets exhibit a higher proportion of hot electrons than higher-coordinated surface sites such as \{111\} facets or the core sites.
The present results thereby demonstrate how hot carriers could be tailored by careful design of atomic-scale structures in nanoscale systems.
\end{abstract}

\maketitle

Plasmon-enhanced technologies enabled by metal \glspl{np} provide promising avenues for harvesting and converting sunlight to chemical energy \cite{AslRaoCha18} and driving photochemical reactions \cite{LinAslBoe15}.
The underlying processes rely on the decay of plasmonic excitations and the subsequent generation of high-energy non-equilibrium electrons and holes \cite{BroHalNor15}.
These electrons and holes are often collectively referred to as \glspl{hc}, but their distributions can vary substantially with time after excitation \cite{SaaAseGar16, LiuZhaLin18}.
\Glspl{hc} generated by plasmon decay can in principle be transferred to a chemically attached acceptor such as a semiconductor or a molecule, a process that is potentially useful for technologies such as photovoltaics \cite{AtwPol10}, photo-detection \cite{KniSobNor11, ChaSchBro14}, photon up-conversion,\cite{NaiWelBri17} and photocatalysis \cite{LinAslBoe15, MukLibLar13, MubLeeSin13, KalAvaChr14, SweZhaZho16}, and possibly relevant for \gls{np} growth processes \cite{ZhaDuCWan16}.

It can be challenging to develop comprehensive understanding of plasmon-generated hot carriers via purely experimental approaches both due to time constraints and the difficulty associated with disentangling different contributions \cite{ZhoSweZha18, SivBarUn19, DubSiv19}.
In this context, complementary theoretical and computational approaches can provide highly valuable insight as they enable scrutinizing the relevant microscopic processes.
The present theoretical understanding of plasmonic \gls{hc} generation is mostly based on flat metal surfaces \cite{SunNarJer14, BerMusNea15} or jellium \glspl{np} neglecting the underlying atomic structure. \cite{ManLiuKul14, SaaAseGar16, BesGov16, YanWanMen16, LiuZhaLin18, DalRanLis18, RanDalLis18, RomHesLis19, RomKahHes20}
While atomic-scale effects in nanoplasmonics in general have been increasingly addressed in recent years, \cite{Zhang2014, KuiSakRos15, Rossi2015Quantized, Marchesin2016, DonLinAik17, SenLinKud19} atomic-scale modeling of plasmonic \gls{hc} generation is only emerging. \cite{MaWanWan15, DouBerSan16, DouBerFra19}
In particular, detailed atomic-scale distributions of plasmon-generated \glspl{hc}, to our knowledge, have not yet been scrutinized.
In the context of photocatalysis especially, detailed understanding of plasmonic \gls{hc} generation at the atomic scale is, however, of paramount importance as chemical reactions take place at this size scale.

In this work, we analyze the effect of local atomic-scale structure on plasmonic \glspl{hc} generation and demonstrate that the distribution of \glspl{hc} after plasmon decay is sensitive to the atomic-scale details.
Using a series of \glspl{np}, we analyze quantitatively the spatial distribution of plasmon-generated \glspl{hc} at the atomic scale with respect to surface orientation as well as for edge and corner sites.
Since the trends are consistent across \glspl{np} of different size and shape, the trends obtained here are expected to be transferable to more general nanoscale structures.
For this study, we have developed a fully atomistic, parameter-free, and generally applicable description of plasmonic \gls{hc} generation based on \gls{ks} \gls{dft} \cite{Hohenberg1964, Kohn1965} and \gls{tddft}. \cite{Runge1984}

\begin{figure*}[t]
    \centering
    \includegraphics[scale=1]{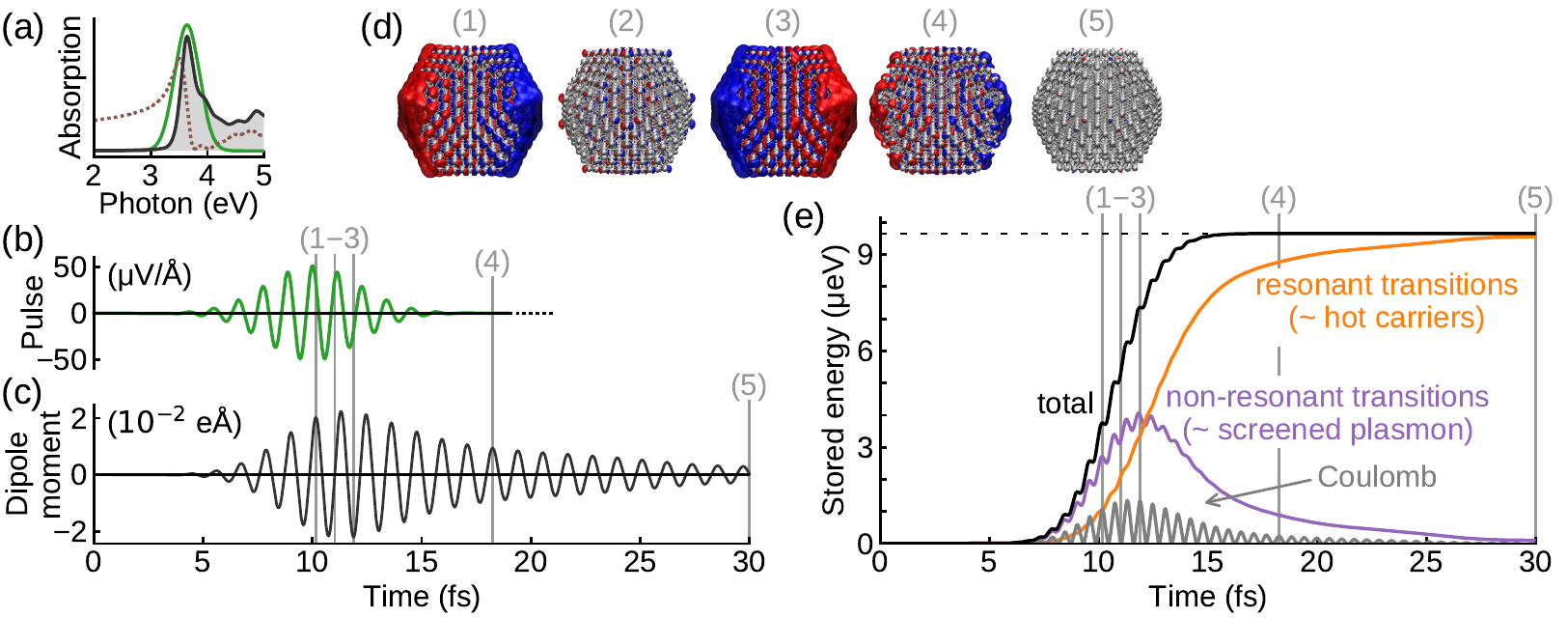}
    \caption{%
        \textbf{Real-time dynamics of a localized surface plasmon in a silver nanoparticle.}
        (a) Photoabsorption spectrum of the \ce{Ag561} \gls{np} (shaded) and the intensity profile of an impacting electric field pulse (green).
        Photoabsorption is determined by the imaginary part of the polarizability $\alpha$, and the corresponding real part $\Re[\alpha]$ is shown as a dotted line.
        (b) Electric field pulse impacting the plasmon resonance of the \gls{np}.
        (c) Time-dependent dipole moment response of the \gls{np}.
        (d) Electron density oscillations in the \gls{np} at selected time instances (red and blue isosurfaces denote density increase and decrease, respectively).
        (e) Time evolution of the energy stored in the excited electronic system.
        The total energy (black) is divided into the energy of non-resonant electron-hole transition contributions constituting screened plasmon excitation (purple) and that of resonant transition contributions constituting mainly hot carriers (orange).
        A part of the plasmon energy is in the form of Coulomb energy (grey).
    }
    \label{fig:energy}
\end{figure*}

\section*{Results and Discussion}

\textbf{Real-Time Dynamics of Localized Surface Plasmon.}
To introduce our approach for modeling plasmonic \gls{hc} generation, we start with a comprehensive characterization of plasmon formation and subsequent dephasing.
We consider an icosahedral \ce{Ag561} silver \gls{np} as an example system with a clear plasmon resonance in the photoabsorption spectrum (\autoref{fig:energy}a) \cite{RosKuiPus17}.
The ground-state electronic structure of the \gls{np} is calculated with \gls{dft} using the Gritsenko--van Leeuwen--van Lenthe--Baerends--solid-correlation (GLLB-sc) \gls{xc} potential \cite{GriLeeLen95, KuiOjaEnk10} for an improved d-band description. \cite{Yan2011First, KuiSakRos15}
The time-dependent response is then calculated with \gls{tddft} using either the \gls{rpa} or the adiabatic GLLB-sc \cite{KuiSakRos15} (see \nameref{sec:methods} for details).

To excite the \gls{lspr} in \gls{np}, we use a monochromatic ultrafast Gaussian light pulse
\begin{align}
    \mathcal{E}(t) = \mathcal{E}_{0} \cos\big(\omega_0 (t - t_0)\big) \exp\big(-(t - t_0)^2/\tau_0^2\big)
    \label{eq:pulse}
\end{align}
that induces real-time dynamics of electrons in the system.
The pulse frequency $\omega_0=\unit[3.6]{eV}$ is tuned to the plasmon resonance, the pulse duration is determined by $\tau_0 = \unit[3]{fs}$,
and the pulse is centered at $t_0 = \unit[10]{fs}$ (\autoref{fig:energy}b).
The pulse strength is weak, $\mathcal{E}_{0}=\unit[51]{\text{\textmu{}V/\angstrom}}$, putting the response in the linear-response regime.
In the frequency space, the pulse is wide enough to cover the whole plasmon resonance (\autoref{fig:energy}a).

The interaction between electrons and light is described in the dipole approximation, within which the light pulse creates a time-dependent external potential $v_{\mathrm{pulse}}(t) = z \mathcal{E}(t)$ that causes the time evolution of the \gls{ks} states $\ket{\psi_n(t)}$ and excitation of the \gls{lspr}.
The light pulse induces a strong dipole-moment response [\autoref{fig:energy}c(1--3)].
The corresponding electron density oscillations [\autoref{fig:energy}d(1--3)] are composed of a surface-to-surface component associated with delocalized valence electrons near the Fermi energy and atom-localized contributions that correspond to screening due to virtual excitations from the d-band. \cite{RosKuiPus17}
As time proceeds to $t\approx \unit[17]{fs}$, the excited electrons start to lose their collective plasmonic motion via a dephasing process commonly referred to as Landau damping. \cite{Yannouleas1992}
As the plasmon dephases, the dipole moment decays [\autoref{fig:energy}c(4--5)] corresponding to vanishing surface-to-surface density oscillation [\autoref{fig:energy}d(4--5)]. \cite{MaWanWan15}
While the density oscillations and dipole moment response gives an illustrative picture of the plasmon formation and decay, they seem not to provide a tractable way for scrutinizing \gls{hc} contributions to the response.
To this end, we consider the time-dependent energy contributions of different \gls{ks} transitions.

\begin{figure*}[t]
    \centering
    \includegraphics[scale=1]{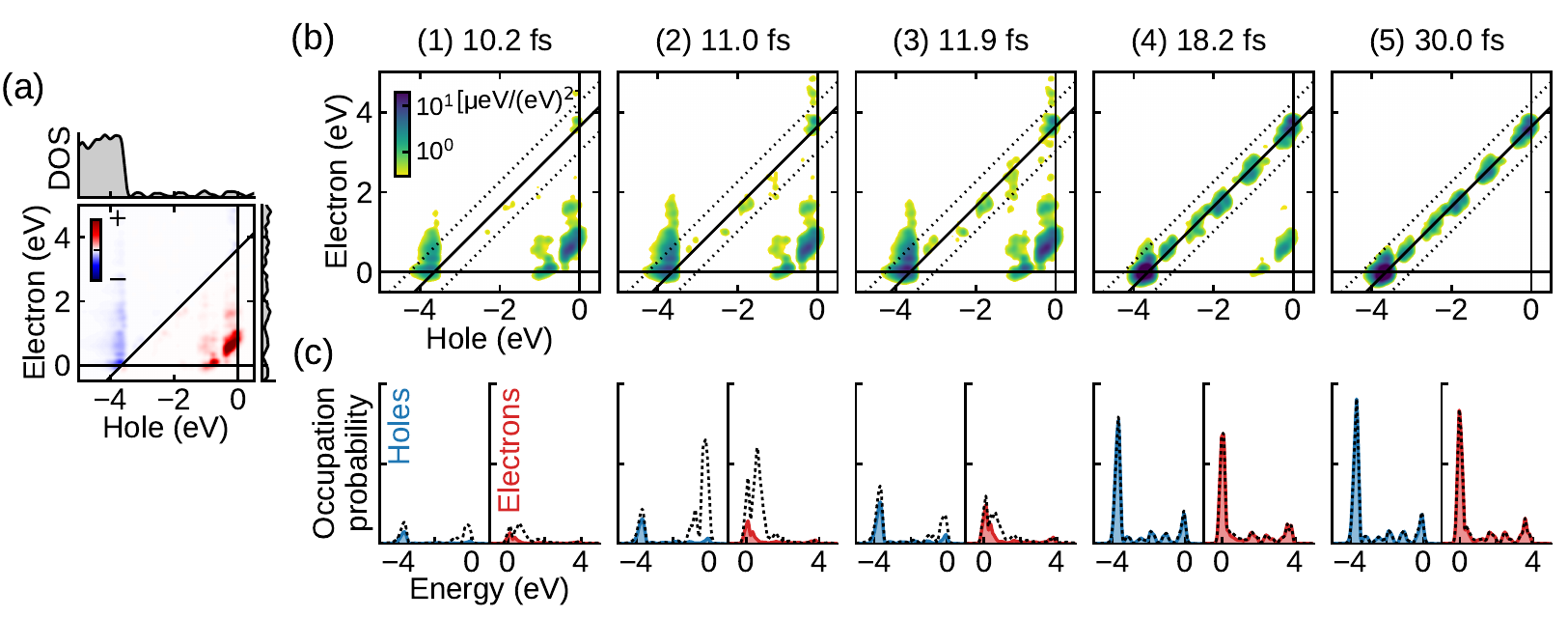}
    \caption{%
        \textbf{Electron-hole transition contributions to plasmon formation and decay.}
        (a) Electron-hole contributions to the photoabsorption at the resonance energy visualized as a transition contribution map (TCM).
        Density of states (DOS) is also shown along the energy axes.
        (b) Electron-hole contributions to the time-dependent electronic energy visualized as TCM on a logarithmic color scale.
        The solid diagonal line corresponds to the transition energies matching with the pulse frequency $\omega_0$ and the dotted diagonal lines are drawn at $\omega_0 \pm 2\sigma$ to indicate the pulse width $\sigma = \sqrt{2}/\tau_0$, defining the boundaries for resonant and non-resonant transitions for \autoref{fig:energy}e.
        (c) Occupation probabilities of hole and electron states.
        Solid blue and red lines denote state occupations from resonant transitions, and dashed lines denote occupations from all transitions (resonant and non-resonant).
        The figure columns (1--5) show panels (b) and (c) for the time instances labeled in \autoref{fig:energy}.
        The color scale and axis limits are the same in all the columns.
    }
    \label{fig:snapshots}
\end{figure*}

\textbf{Time-Dependent Energy Contributions.}
Since the pulse is tuned to the \gls{lspr}, the electronic system absorbs energy from the incident light and remains in an excited state after the pulse has vanished.
To analyze the distribution of this energy, we first consider the total time-dependent energy of the system given by
\begin{align}
    E_\mathrm{tot}(t) = E^{(0)}_\mathrm{tot} + \Delta E(t) + E_\mathrm{pulse}(t),
\end{align}
where $E^{(0)}_\mathrm{tot}$ is the ground-state energy,
$\Delta E(t)$ is the time-dependent energy stored in the excited state (\autoref{fig:energy}e, black line), and
$E_\mathrm{pulse}(t) = - \mu(t) \mathcal{E}(t)$ is the potential energy of the system under the external electric field.

The incident light pulse pumps energy into the system, \ie, it does work on the system as $\Delta \dot E(t) = \delta\dot\mu(t) \mathcal{E}(t)$, where dots indicate time derivatives and $\delta\mu(t) = \mu(t) - \mu^{(0)}$ is the induced dipole moment.
Thus, the total accumulated electronic energy can be written as
\begin{align}
    \Delta E(t) = \int_{0}^{t} \delta\dot\mu(\tau) \mathcal{E}(\tau) \dee \tau.
\end{align}
The electronic energy increases through absorption in a step-wise manner following the pulse intensity (\autoref{fig:energy}e, $t\approx 5 \dots \unit[15]{fs}$).
After the pulse has ended ($t \gtrsim \unit[15]{fs}$), the absorbed energy remains in the system and the total energy has attained a new constant value given by the photoabsorption cross section [\autoref{fig:energy}a; Eq.~\eqref{eq:totabs} in \nameref{sec:methods}].

While the total energy remains constant, the electronic energy does not stay equally distributed among the electron-hole transitions $i \to a$ excited by the light pulse.
To quantify this effect, we consider the decomposition of the energy in electron-hole transition contributions.
This decomposition is based on the linear response of the \gls{ks} density matrix, $\delta\rho_{ia}(t)$, expressed in the basis of the eigenstates $(\epsilon_n, \psi^{(0)}_n)$ of the ground-state Hamiltonian.
The electron-hole decomposition of energy is (see Supplementary Note~S1 for derivation)
\begin{align}
    \Delta E(t) = \sum_{ia}^{f_i > f_a} \omega_{ia} P_{ia}(t) + E^{\mathrm{C}}_{ia}(t) ,
    \label{eq:E}
\end{align}
where the sum is restricted by ground-state occupation numbers $f_n$ so that the indices $i$ and $a$ correspond to the created hole and electron states, respectively.
Here, $\omega_{ia} = \epsilon_{a} - \epsilon_{i}$ is the electron-hole transition energy (the \gls{ks} eigenvalue difference),
$P_{ia}(t)$ is the transition probability defined as
\begin{align}
    P_{ia}(t) = \left|\frac{\delta\rho_{ia}(t)}{\sqrt{f_{i}-f_{a}}}\right|^2 ,
    \label{eq:tprob}
\end{align}
and $E^{\mathrm{C}}_{ia}(t)$ is the Coulomb energy as obtained from the Hartree--\gls{xc} kernel (defined in Supplementary Note~S1).

Plasmon formation and dephasing can be scrutinized by considering the energy stored in the electronic system in terms of the electron-hole transition energy $\omega_{ia}$ with respect to the pulse frequency $\omega_0$ (Supplementary Fig.~S1).
The plasmon is formed by constructive coupling of low-energy transitions [$\omega_{ia} \lesssim\!\unit[2]{eV}$; see \autoref{fig:snapshots}a and time instances (1--3) in \autoref{fig:snapshots}b]. \cite{Yannouleas1992, Bernadotte2013, RosKuiPus17}
Simultaneously, high-energy virtual d-electron transitions ($\omega_{ia} \gtrsim\!\unit[4]{eV}$) screen the plasmonic density oscillation, decreasing the total induced field.
The energy stored in the screening is similar to the energy stored in the polarization of a dielectric in general; \ie\ as the total field strength decreases, energy is instead stored in the polarization of the d-electron states as in $\omega_{ia} P_{ia}(t)$ in Eq.~\eqref{eq:E}.
The non-resonant low and high-energy transitions carry most of the energy during plasmon excitation (\autoref{fig:energy}e, purple line).
As the plasmon dephases, the absorbed energy is redistributed to electron-hole transitions that are resonant with the pulse [\autoref{fig:energy}e, orange line; corresponding to the diagonal in the transition contribution maps in \autoref{fig:snapshots}b; see time instances (4--5)].
After dephasing, the energy remains almost exclusively stored in these transitions constituting the plasmon-generated \glspl{hc}.
The transitions comprising the plasmon are active in photoabsorption (\autoref{fig:snapshots}a), and hot holes and electrons are generated through plasmon decay, instead of the \gls{hc} transitions absorbing the light directly (shown in detail in Supplementary Fig.~S2).

At longer time scales, the electronic system would dissipate the absorbed energy to the environment via radiation, atomic motion, or other processes, but such decay pathways are not included within the description used here.
Thus, in the present picture there appears no significant dynamics in \ce{Ag561} at time scales beyond $t \gtrsim \unit[30]{fs}$ after the fast dephasing of the \gls{lspr} through Landau damping, which takes place due to the presence of multiple excitation eigenstates forming the broadened plasmon peak in the photoabsorption spectrum \cite{Yannouleas1992} (Supplementary Fig.~S3).
However, the dynamics can be very different in small clusters with discrete excitation spectrum.
In contrast to a single broad plasmon peak in \ce{Ag561}, for example in the \ce{Ag55} cluster individual electron-hole transitions couple strongly to the plasmon, \cite{RosKuiPus17} a process that is referred to as plasmon fragmentation. \cite{Yannouleas1989, Yannouleas1991}
Correspondingly, the time-domain response exhibits Rabi oscillations and energy transfer back from the resonant transitions to the plasmon \cite{MaWanWan15, YouRamSei18} (Supplementary Fig.~S4).

Since the coupling of transitions via Coulomb interaction is recognized as an essential characteristic of plasmonic excitations, \cite{Yannouleas1992, Bernadotte2013, Zhang2017, RosKuiPus17} it is instructive to consider the Coulomb energy $E_\mathrm{C}(t) = \sum_{ia} E^\mathrm{C}_{ia}(t)$. 
This energy exhibits strong oscillations (\autoref{fig:energy}e, grey line) analogous to the dipole moment (\autoref{fig:energy}c) as only the electron density oscillation contributes to the Coulomb energy.
At the maxima of the surface-to-surface density oscillation [time instances (1) and (3) in Figs.~\ref{fig:energy}d--e], the Coulomb contribution is a significant part of the plasmon energy, but at the minima in between [\eg, time instance (2)] the Coulomb energy is vanishing as the electronic energy is stored in the electron current flowing through the particle (Supplementary Fig.~S5).

\textbf{Temporal Evolution of Hot-Carrier Distributions.}
Having established a real-time picture of plasmon formation and decay, we are in the position to analyze the distributions of electrons and holes during the process.
The probabilities for creating a hole in an initially occupied state $i$ or an electron in an initially unoccupied state $a$ are given directly by the transition probability of Eq.~\eqref{eq:tprob} as
\begin{align}
    P^\mathrm{h}_{i}(t) = \sum_{a}^{f_i > f_a} P_{ia}(t) \quad\text{ and }\quad
    P^\mathrm{e}_{a}(t) = \sum_{i}^{f_i > f_a} P_{ia}(t),
\label{eq:occ}
\end{align}
respectively.
$P^\mathrm{h}_{i}$ and $P^\mathrm{e}_{a}$ determine exactly the diagonal elements of the second order response of the density matrix (Supplementary Note~S1); in other words, they correspond to the increase of the occupation of the initially unoccupied state $a$ and the decrease of the occupation of the initially occupied state $i$, respectively.

The occupation probabilities given by Eq.~\eqref{eq:occ} show strong oscillations during the time evolution [\autoref{fig:snapshots}c(1--3); dashed lines].
These oscillations are explained by the oscillation of the Coulomb energy.
As the Coulomb energy contribution is carried mainly by non-resonant transitions (Supplementary Fig.~S6), the occupation probabilities of the electron and hole states contributing to these non-resonant transitions oscillate analogously to the Coulomb energy.
The oscillations are especially visible in the occupations of electron and hole states that form the plasmon, \cite{DouBerFra19} \ie, the states near the Fermi energy, often referred to as Drude carriers \cite{HarBesJoh17}.
The oscillatory population and depopulation of these states indicate that they would not likely be individually separable while they are a part of the plasmon as the Coulomb interaction is an essential part of the excitation itself \cite{AizBalBau19}.

The contributions of the resonant transitions to the Coulomb energy are relatively small (Supplementary Fig.~S6) and the occupations of the corresponding electron and hole states grow steadily as the plasmon decays (\autoref{fig:snapshots}c; solid lines).
At the end of the dynamic evolution studied here [\autoref{fig:snapshots}c(5)], electrons and holes are still coupled in the form of electron-hole transitions, and the distributions at $t=\unit[30]{fs}$ can be considered as the initial non-thermal \gls{hc} distributions.
After their generation these carriers would separate and interact via electron--electron and electron--phonon scattering processes \cite{BerMusNea15, BroHalNor15, LiuZhaLin18, Khu20} that are not captured in the present description.
The slight asymmetry between the hole and electron distributions is caused by a non-zero width of the pulse in the frequency space (\autoref{fig:energy}a).

\begin{figure*}[t!]
    \centering
    \includegraphics[scale=1]{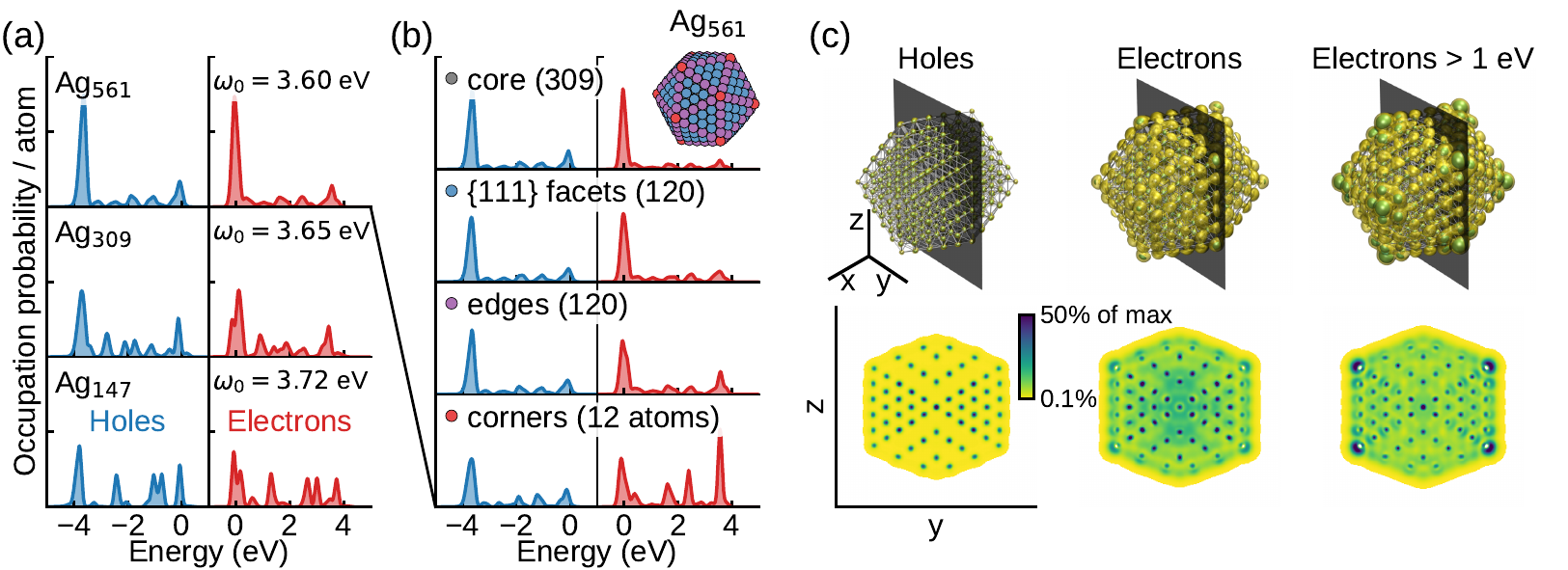}
    \caption{%
        \textbf{Hot-carrier distributions after plasmon decay.}
        (a) Occupation probabilities of hole and electron states in icosahedral silver \glspl{np} of 147--561 atoms.
        (b) Occupation probabilities at different atomic sites of \ce{Ag561}.
        All the panels use the same axis limits (normalized by the number of atoms).
        (c) Spatial density profiles of all induced holes and electrons and induced electrons with energy of more than \unit[1]{eV} in \ce{Ag561}.
        Plots show isosurfaces corresponding to 10\% and 20\% of maximum value, respectively, and slices are taken through the center of the \gls{np}.
    }
    \label{fig:hcdist}
\end{figure*}

\textbf{Energetic and Atomic-Scale Spatial Distributions of Hot Carriers.}
We now analyze the distribution of plasmon-generated hot carriers and the impact of local atomic-scale structure.
We start by considering the series of icosahedral silver \glspl{np} \ce{Ag147}, \ce{Ag309}, and \ce{Ag561}, the photoabsorption properties of which we have described in detail in earlier work \cite{KuiSakRos15, RosKuiPus17} (see Supplementary Fig.~S7 for photoabsorption spectra and densities of states).
The light pulse is tuned to the plasmon resonance of the \glspl{np} and the initial \gls{hc} distributions are analyzed after the plasmon has dephased at time $t=\unit[30]{fs}$.
The \gls{hc} distributions show a pronounced dependence on \gls{np} size (\autoref{fig:hcdist}a) and local structure (\autoref{fig:hcdist}b--c) as discussed in the following.

As particle size increases, the \gls{hc} distributions are increasingly dominated by interband d-electron transitions \cite{RomKahHes20} (hole $\sim\!\!\unit[-4]{eV}$ $\to$ electron $\sim\!\!\unit[0]{eV}$) converging toward the distributions obtained for flat surfaces. \cite{SunNarJer14, BerMusNea15}
In contrast to extended systems, geometry confinement effects are significant for plasmonic \gls{hc} generation in nanoscale systems. \cite{BroSunNar16}
Due to the broken crystal symmetry, additional ``intraband'' transitions are available for plasmonic \gls{hc} generation in \glspl{np}, which results in the population of higher-energy electron and hole states (\autoref{fig:hcdist}a; electron states of $> \unit[0.5]{eV}$, hole states of $>\unit[-3.5]{eV}$).
The relative contribution of these sp-states is most pronounced in the smallest \glspl{np} \cite{RomKahHes20} (\ce{Ag147}, \ce{Ag309}) but they are non-negligible also in \ce{Ag561}.
Similar size-dependent trends are also present in silver \glspl{np} of other shapes, while the detailed relative contributions of different transitions vary (Supplementary Fig.~S8).

The calculated probability distributions of plasmon-generated electrons and holes (see \nameref{sec:methods}) exhibit strong spatial variance (shown for the icosahedral \ce{Ag561} \gls{np} in \autoref{fig:hcdist}b--c):
Holes are localized at atomic sites throughout the particle, which is expected as the majority of holes originates from the atom-localized d-states.
As a result, their energy distribution is very similar for core and surface sites.
Hot electrons, on the other hand, are more delocalized and reside to larger extent in the surface region.
The surface contribution is even more pronounced for higher-energy hot electrons ($>\unit[1]{eV}$ electrons in \autoref{fig:hcdist}c).
The probability density of these hot electrons is strongly enhanced especially at low-coordinated edge and corner sites compared to sites in the core and on flat surfaces (\autoref{fig:hcdist}b).
The energetic distributions of plasmon-generated holes and electrons are not necessarily symmetric when projected onto a particular site (\autoref{fig:hcdist}b; Supplementary Fig.~S8).
This asymmetry is especially pronounced for the corner site in \ce{Ag561}, which reflects the fact that the hot-electron density at corner sites originates likely from throughout the particle due to the uniformity of the hole density.

\begin{figure*}[t]
    \centering
    \includegraphics[width=\textwidth]{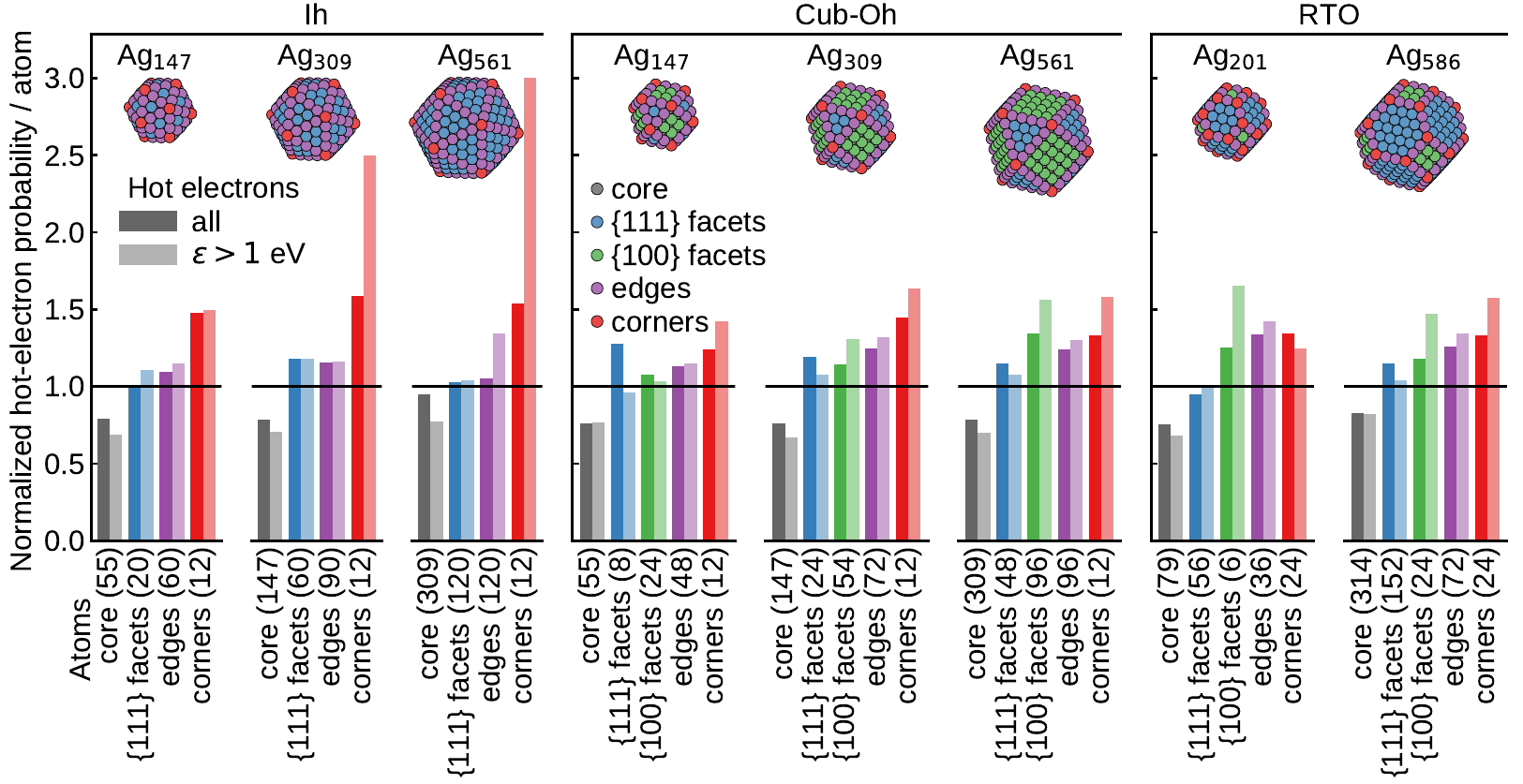}
    \caption{%
        \textbf{Atomic-scale distributions of hot electrons in silver nanoparticles.}
        Spatial distribution of hot electrons generated on different atomic sites in icosahedral (Ih), cuboctahedral (Cub-Oh), and regularly truncated octahedral (RTO) \glspl{np}.
        Sites with lower coordination exhibit a higher proportion of hot electrons than core sites.
        A spatially uniform distribution corresponds to a normalized probability of unity.
        The insets show the atomic structures with the different atomic sites colored.
    }
    \label{fig:heatom}
\end{figure*}

A more quantitative view is obtained by considering the total per-atom occupation probability of hot electrons at a particular atomic site in comparison to the total per-atom occupation probability throughout the system (\autoref{fig:heatom}).
This data suggests that hot electrons with more than \unit[1]{eV} are almost three times more likely to be found in the vicinity of a corner site in the icosahedral \ce{Ag561} \gls{np} than a uniform spatial distribution would correspond to.
We emphasize that these are per-atom considerations, that is, as the 12 corner atoms constitute only around 2\%{} of the atoms in the particle, it is expected that about 6\%{} of the electrons with more than \unit[1]{eV} would be generated in the vicinity of the corner atoms.
Overall, hot electrons with more than \unit[1]{eV} constitute 30 to 60\%{} of all hot electrons depending on system (Supplementary Fig.~S8).
The absolute total number of \glspl{hc} generated depends on the light energy that is absorbed, which is in turn determined by photoabsorption cross-section and light intensity.

Similarly to icosahedral shapes, hot electrons in the cuboctahedral and regularly truncated octahedral \glspl{np} are more likely to be generated at a surface site than at a core site (\autoref{fig:heatom}).
The preference for surface sites is even more pronounced for hot electrons with more than \unit[1]{eV}.
The corner sites of the cuboctahedral and regularly truncated octahedral \glspl{np}, however, do not show as enhanced distribution of hot electrons as those sites of the icosahedral \ce{Ag561} particle.
This further underlines the sensitivity of \gls{hc} generation to atomic-scale details and the exact electronic structure of the \gls{np} and site(s) in question.

As a general trend, lower-coordinated sites seems to exhibit an enhancement of hot electrons compared to higher-coordinated sites.
This trend is observed for corner and edge sites, but the data for cuboctahedral and regularly truncated octahedral \glspl{np} suggests also that more hot electrons are generated on the lower-coordinated \{100\} surface than on the \{111\} surface.
In contrast to strong spatial variation of hot electrons, plasmon-generated holes do not show strong spatial dependence in the considered \glspl{np} (Supplementary Fig.~S9).

For practical utilization, \glspl{hc} need to be transferred to the environment.
The \gls{hc} distributions obtained in the present work can be considered as the initial non-thermal \gls{hc} distributions before any electron--electron scattering \cite{BerMusNea15, Khu20} has taken place.
Thus, these \gls{hc} distributions could serve as an initial condition for subsequent dynamics \cite{BroHalNor15, LiuZhaLin18} that is not included in the present approach. 
In general, \gls{hc} transfer to environment can occur \emph{indirectly}, \ie\ carriers are first produced in the metal and subsequently transferred to the acceptor, or \emph{directly}, \ie, plasmon dephasing leads directly to the injection of \glspl{hc} in empty acceptor (or occupied donor) states. \cite{LinAslBoe15, NarSunAtw16, ChrMos17, WuCheMcB15, KalChr15, BoeCamMor16, BoeAslLin16, TanArgRen17, TanDaiZha18, LiDiSMur17, FoeJopKae17}
However, both experiments \cite{RatDunVur17, TanLiuDai17, BroSunNar17} and calculations \cite{BroSunNar16, BroSunNar17, BesKonWan17, Khu20} indicate that \glspl{hc} generated in the metal can quickly relax through electron--electron scattering, which renders the indirect pathway inefficient.
The direct-transfer process, on the other hand, presents an opportunity to obtain more efficient plasmonic \gls{hc} devices \cite{ChrMos17, FoeJopKae17, KalChr15}.
The prevalence of hot electrons on lower-coordinated surface sites described in the present work seems favorable for their utility through the direct transfer processes.
It is, however, crucial to also consider the hybridization of the surface electronic states with acceptor states, where the latter can originate, \eg, from an adsorbed molecule \cite{KumRosKui19} or a semiconductor \cite{KumRosMar19, MaGao19}.
To maximize the efficiency for direct excitation transfer the emitting and receiving states should be energetically aligned and spatially overlapping.
In addition, hot-electron generation can be affected by the dielectric environment, \eg, by red-shifting the plasmon resonance so that d-band electrons are not excited.
The framework presented here enables analysis and quantification of these aspects at the atomic scale with material specificity and without resorting to empirical parameters.

\section*{Conclusions}

In conclusion, we presented a comprehensive first-principles account of the real-time dynamics of plasmon formation and dephasing into \glspl{hc} and analyzed quantitatively the impact of atomic-scale structure on the \gls{hc} generation.
Our results on silver \glspl{np} indicate that lower-coordinated surface sites exhibit a larger proportion of \emph{hot electrons}, especially those with higher energy, than the bulk of the nanoparticle or higher-coordinated surface sites.
In contrast, the distribution of \emph{hot holes} is relatively homogeneous within each considered \gls{np}.
These features can be traced to the electronic structure as hole and electron states exhibit localized d and delocalized sp-type character, respectively.
We would therefore expect the present insight to be qualitatively transferable to other late transition metals that exhibit similar electronic structure.
The observed trends are present in \glspl{np} of different shapes and sizes with varying prevalence.
We therefore envision that the obtained atomic-scale insights could be applicable to nanoscale structures in general.
First-principles predictions of plasmonic \gls{hc} generation, as presented here, can thereby facilitate tuning and optimizing photocatalytic systems down to the atomic scale by, \eg, enabling identification of promising surface-acceptor combinations.

\section*{Methods}
\label{sec:methods}

\textbf{Computational Details.}
The ground-state electronic structures were calculated with \gls{ks}-\gls{dft} \cite{Hohenberg1964, Kohn1965}
using the GLLB-sc exchange-correlation potential. \cite{GriLeeLen95, KuiOjaEnk10}
The time-domain responses were calculated with \gls{tddft} \cite{Runge1984} starting from the ground state.
The dynamical response was described with \gls{rpa} for the data in Figs.~\ref{fig:energy}--\ref{fig:snapshots} and in Supplementary Figs.~S1--S6, while all the other data was calculated with the adiabatic GLLB-sc. \cite{KuiSakRos15}
The two response kernels yield very similar results (Supplementary Fig.~S10), but the GLLB-sc potential is not suitable for obtaining the total energies.

All the calculations are carried out with the open-source GPAW code package \cite{Enkovaara2010} using localized basis sets \cite{Larsen2009} and the real-time propagation \gls{tddft} implementation. \cite{KuiSakRos15}
We used 11-electron projector augmented-wave \cite{Blochl1994} setups for Ag, treating the remaining electrons as frozen core.
We used an extended ``p-valence'' basis set that includes diffuse 5p functions, which are important for describing plasmon resonances \cite{Rossi2015Nanoplasmonics}.
The basis set is similar to the ones used in Refs.~\onlinecite{KuiSakRos15} and \onlinecite{RosKuiPus17}.

The photoabsorption spectra were calculated using the $\delta$-kick technique \cite{Yabana1996} yielding linear impulse response.
The photoabsorption of icosahedral particles is isotropic and the electric field was aligned along the $x$ direction.
The resulting \glspl{hc} do not exhibit a strong variation between different sites (\autoref{fig:hcdist}c).
For the time propagation, we used a time step of \unit[10]{as} and total propagation time of at least \unit[30]{fs}.
The spectra were broadened using Gaussian damping with $\sigma = \unit[0.07]{eV}$ corresponding to a full width at half-maximum of \unit[0.16]{eV}.
The real-time response to a pulse was calculated as a post-processing step through convolution as described below in detail.
In the convolution Fourier transforms or time-domain response there is no artificial damping.

A grid spacing parameter of $h=\unit[0.3]{\angstrom}$ was chosen to represent densities and potentials and the nanoparticles were surrounded by a vacuum region of at least \unit[6]{\angstrom}.
The Hartree potential was evaluated with a Poisson solver using the monopole and dipole corrections for the potential.
Fermi-Dirac smearing was applied to the occupation numbers to facilitate convergence.
The \gls{ks} electron-hole basis included electron-hole pairs with occupation number difference $f_i - f_a \geq 10^{-3}$.

Before the response calculations, all geometries were relaxed using the BFGS optimizer in the open-source ASE package. \cite{Larsen2017}
The relaxation calculations used the Perdew--Burke--Ernzerhof (PBE) \cite{Perdew1996} functional, double-$\zeta$ polarized (dzp) basis sets, and $h=\unit[0.2]{\angstrom}$.

\textbf{Pulse Response from Convolution.}
The photoabsorption spectrum can be efficiently calculated from real-time propagation using the $\delta$-kick perturbation \cite{Yabana1996} as in the linear-response regime all the frequencies are independent of each other.
We utilize this property in the present work for calculating the linear response of the density matrix to the Gaussian pulse of Eq.~\eqref{eq:pulse} as a post-processing step.
First, the time-propagation is carried out for perturbation $v_\mathrm{kick}(t) = z K_0 \delta(t)$ yielding the impulse response of the system and the corresponding time-dependent density matrix $\delta\rho^\mathrm{kick}_{ia}(t)$.
Then, in the linear-response regime, the response to the pulse of Eq.~\eqref{eq:pulse} is obtained as a convolution
\begin{align}
    \delta\rho_{ia}(t) = \frac{1}{K_0} \int_{0}^{\infty} \delta\rho^\mathrm{kick}_{ia}(\tau) \mathcal{E}(t - \tau) \dee \tau,
\end{align}
which can be very efficiently calculated in frequency space by employing the convolution theorem and inverse Fourier transform
\begin{align}
    \delta\rho_{ia}(t) = \frac{1}{2 \pi K_0} \int_{-\infty}^{\infty}  \delta\rho^\mathrm{kick}_{ia}(\omega) \mathcal{E}(\omega) e^{-i \omega t} \dee \omega ,
    \label{eq:conv:f}
\end{align}
where $\delta\rho^\mathrm{kick}_{ia}(\omega)$ and $\mathcal{E}(\omega)$ are Fourier transforms of the respective time-domain quantities.
Here, $\delta\rho^\mathrm{kick}_{ia}(\omega)$ can be efficiently calculated from the impulse response by using the computational framework developed in Ref.~\onlinecite{RosKuiPus17}.
The time derivatives required for calculating the energy (Supplementary Note~S4) are obtained similarly as
\begin{align}
    \delta\dot\rho_{ia}(t) &= - \frac{i}{2 \pi K_0} \int_{-\infty}^{\infty} \omega \delta\rho^\mathrm{kick}_{ia}(\omega) \mathcal{E}(\omega) e^{-i \omega t} \dee \omega, \\
    \delta\ddot\rho_{ia}(t) &= - \frac{1}{2 \pi K_0} \int_{-\infty}^{\infty} \omega^2 \delta\rho^\mathrm{kick}_{ia}(\omega) \mathcal{E}(\omega) e^{-i \omega t} \dee \omega .
\end{align}
In practice $\mathcal{E}(\omega)$ is non-vanishing only on a finite frequency interval (see, \eg, \autoref{fig:energy}a), which narrows the integration limits.

It is emphasized that the time-dependent density matrix $\delta\rho_{ia}(t)$ is a complex quantity in time domain, so in practical calculations it is convenient to carry out Fourier transform for the real $\Re\delta\rho_{ia}(t)$ and imaginary $\Im\delta\rho_{ia}(t)$ parts separately to utilize the properties of Fourier transform of real quantities.

We also note in passing that the impulse response $\delta\rho^\mathrm{kick}_{ia}(\omega)$ can be equivalently calculated from the Casida linear-response frequency-space formalism \cite{Casida1995, RosKuiPus17}.
Hence, the linear real-time response to any pulse can also be calculated from the Casida solutions by using the convolution of Eq.~\eqref{eq:conv:f}.

\textbf{Total Absorbed Energy.}
By invoking Fourier transform, the total absorbed energy after the pulse has vanished is obtained as
\begin{align}
    \int_{0}^{\infty} \delta\dot\mu(t) \mathcal{E}(t) \dee t
    = \frac{1}{2} \int_{0}^{\infty} S(\omega) \left| \mathcal{E}(\omega) \right|^2 \dee \omega ,
    \label{eq:totabs}
\end{align}
where $S(\omega)=\frac{2\omega}{\pi} \Im[\alpha(\omega)]$ is the dipole strength function, which equals the photoabsorption cross section safe for a constant multiplier.

\textbf{Hot-Carrier Distributions.}
The hot-electron energy distributions corresponding to the state occupation probabilities $P^\mathrm{e}_{a}$ of Eq.~\eqref{eq:occ} are obtained as (time-dependence is not explicitly marked)
\begin{align}
    P_\mathrm{e}(\epsilon) &= \sum_{a} P^\mathrm{e}_{a} \delta(\epsilon - \epsilon_{a})
    = \frac{1}{2} \sum_{ia}^{f_i > f_a} (q^2_{ia} + p^2_{ia}) \delta(\epsilon - \epsilon_{a}) ,
    \label{eq:hcdist}
\end{align}
where
$q_{ia}(t) = 2\Re\delta\rho_{ia}(t) / \sqrt{2(f_i - f_a)}$
and
$p_{ia}(t) = -2\Im\delta\rho_{ia}(t) / \sqrt{2(f_i - f_a)}$,
\ie, they correspond to the real and imaginary parts of $\delta\rho_{ia}$ (see Supplementary Note~S2 for details).
For visualization purposes, Gaussian smoothing (convolution) is applied with respect to the $\epsilon$ axis.

The spatial probability density of hot electrons is obtained by using the full electron-electron part of the second-order density matrix as (see Supplementary Note~S1 and note that only the real part contributes due to the hermiticity of the density matrix)
\begin{align}
    P_\mathrm{e}(\V r) &= \frac{1}{2} \sum_{iaa'}^{\substack{f_i > f_a\\f_a = f_{a'}}} (q_{ia} q_{ia'} + p_{ia} p_{ia'}) \psi^{(0)}_{a}(\V r) \psi^{(0)}_{a'}(\V r) .
    \label{eq:hcdns}
\end{align}
The diagonal and degenerate states dominate the spatial density contributions, which allows us to define a spatio-energetic distribution
\begin{align}
    P_\mathrm{e}(\epsilon, \V r) &= \frac{1}{2} \sum_{iaa'}^{\substack{f_i > f_a\\\epsilon_a = \epsilon_{a'}}} (q_{ia} q_{ia'} + p_{ia} p_{ia'}) \psi^{(0)}_{a}(\V r) \psi^{(0)}_{a'}(\V r) \delta(\epsilon - \epsilon_a),
    \label{eq:hcdistdns}
\end{align}
which is used to calculate the spatial density of hot electrons with \eg, more than \unit[1]{eV} as $P_\mathrm{e}^{>\unit[1]{eV}}(\V r) = \int_{\unit[1]{eV}}^{\infty} P_\mathrm{e}(\epsilon, \V r) \dee \epsilon$, and the energy distribution of hot electrons in a spatial volume $V$ as
$P_\mathrm{e}^{V}(\epsilon) = \int_{V} P_\mathrm{e}(\epsilon, \V r) \dee \V r$.
The distribution at a specific atomic site (\eg, corner atoms) is obtained by integration over the Voronoi cell associated with the site.

The spatial and energetic distributions of hot holes are calculated analogously to the electrons.

\textbf{Software Used.}
The GPAW package \cite{Mortensen2005, Enkovaara2010} with \gls{lcao} mode \cite{Larsen2009} was used for \gls{dft} calculations.
The real-time propagation \gls{lcao}-\gls{tddft} implementation in GPAW \cite{KuiSakRos15} was used for the \gls{tddft} calculations.
Density-matrix-based analysis tools in frequency space \cite{RosKuiPus17} and in real time (present work) were used for analysis.
The ASE library \cite{Larsen2017} was used for constructing atomic structures and geometry relaxation.
The NumPy \cite{Walt2011} and Matplotlib \cite{Hunter2007} Python packages and the VMD software \cite{Humphrey1996, Stone1998} were used for processing and plotting data.

\section*{Data and Code Availability}

Data and code supporting the findings of this study are available at \href{https://doi.org/10.5281/zenodo.3927527}{doi:10.5281/zenodo.3927527}.

\begin{acknowledgments}
We acknowledge financial support from the Knut and Alice Wallenberg Foundation (2014.0226, 2015.0055), the Swedish Research Council (2015-04153), and the Swedish Foundation for Strategic Research (RMA15-0052).
T.P.R.\ acknowledges support from the European Union's Horizon 2020 research and innovation programme under the Marie Sk{\l}odowska-Curie grant agreement No~838996 and also thanks the Adlerbert Research Foundation and the Wilhelm and Martina Lundgren Foundation for support.
M.K.\ acknowledges funding from Academy of Finland under grant No~295602.
We acknowledge generous computational resources provided by the Swedish National Infrastructure for Computing (SNIC) at PDC (Stockholm), NSC (Link\"{o}ping), and C3SE (Gothenburg) as well as by the CSC -- IT Center for Science (Finland).
\end{acknowledgments}

\bibliography{references}

\end{document}